\documentclass[aip,graphicx, reprint]{revtex4-1}

\draft 
\usepackage{graphicx}
\usepackage{placeins}
\usepackage{color}
\usepackage{soul}

\begin{document}


\title{Imaging the Photochemistry of Cyclobutanone using Ultrafast Electron Diffraction: Experimental Results} 

\newcommand{\PULSE}{Stanford PULSE Institute, SLAC National Accelerator Laboratory, 2575 Sand Hill Road, Menlo Park, CA 94025, USA.}
\newcommand{\Edinburgh}{EaStCHEM School of Chemistry, University of Edinburgh, Edinburgh EH9 3FJ, U.K.}
\newcommand{\LCLS}{Linac Coherent Light Source, SLAC National Accelerator Laboratory, 2575 Sand Hill Road, Menlo Park, CA 94025, USA.}
\newcommand{\Bristol}{School of Chemistry, University of Bristol, Bristol BS8 1TS, U.K.}
\newcommand{\Nebraska}{Department of Physics and Astronomy, University of Nebraska -Lincoln, Lincoln, NE 68588, USA}
\newcommand{\Brown}{Department of Chemistry, Brown University, Providence, RI 02912, USA.}
\newcommand{\KState}{Department of Physics, Kansas State University, Manhattan, KS 66506, USA.}
\newcommand{\APStanford}{Applied Physics Department, Stanford University, 348 Via Pueblo, Stanford, CA 94305, USA.}
\newcommand{\EuXFEL}{European XFEL, Holzkoppel 4, 22869 Schenefeld, Germany}
\newcommand{\Davis}{Department of Chemistry, University of California, Davis, One Shields Avenue, Davis, CA 95616, USA.}

\author{A. E. Green}
\affiliation{\PULSE}
\affiliation{\Edinburgh}
\affiliation{\EuXFEL}

\author{Y. Liu}
\affiliation{\LCLS}

\author{F. Allum}
\affiliation{\PULSE}

\author{M. Gra\ss l}
\affiliation{\PULSE}
\affiliation{\LCLS}

\author{P. Lenzen}
\affiliation{\PULSE}

\author{M. N. R. Ashfold}
\affiliation{\Bristol}

\author{S. Bhattacharyya}
\affiliation{\LCLS}

\author{X. Cheng}
\affiliation{\LCLS}

\author{M. Centurion}
\affiliation{\Nebraska}

\author{S. W. Crane}
\affiliation{\Brown}

\author{R. Forbes}
\affiliation{\LCLS}
\affiliation{\Davis}

\author{N. A. Goff}
\affiliation{\Brown}

\author{L. Huang}
\affiliation{\Brown}

\author{B. Kaufman}
\affiliation{\LCLS}

\author{M. F. Kling}
\affiliation{\PULSE}
\affiliation{\LCLS}
\affiliation{\APStanford}

\author{P. L. Kramer}
\affiliation{\LCLS}

\author{H. V. S. Lam}
\affiliation{\KState}

\author{K. A. Larsen}
\affiliation{\LCLS}

\author{R. Lemons}
\affiliation{\LCLS}

\author{M.-F. Lin}
\affiliation{\LCLS}

\author{A. J. Orr-Ewing}
\affiliation{\Bristol}

\author{D. Rolles}
\affiliation{\KState}

\author{A. Rudenko}
\affiliation{\KState}

\author{S. K. Saha}
\affiliation{\KState}

\author{J. Searles}
\affiliation{\KState}

\author{X. Shen}
\affiliation{\LCLS}

\author{S. Weathersby}
\affiliation{\LCLS}

\author{P. M. Weber}
\affiliation{\Brown}

\author{H. Zhao}
\affiliation{\Nebraska}

\author{T. J. A. Wolf}
\email[]{thomas.wolf@slac.stanford.edu}
\affiliation{\PULSE}
\affiliation{\LCLS}



\date{\today}

\begin{abstract}
We investigated the ultrafast structural dynamics of cyclobutanone following photoexcitation at $\lambda=200$ nm using gas-phase megaelectronvolt ultrafast electron diffraction. Our investigation complements the simulation studies of the same process within this special issue. It provides information about both electronic state population and structural dynamics through well-separable inelastic and elastic electron scattering signatures. We observe the depopulation of the photoexcited S$_2$  state of cyclobutanone with n3s Rydberg character through its inelastic electron scattering signature with a time constant of $(0.29 \pm 0.2)$ ps towards the S$_1$ state. The S$_1$ state population undergoes ring-opening via a Norrish Type-I reaction, likely while passing through a conical intersection with S$_0$. The corresponding structural changes can be tracked by elastic electron scattering signatures. These changes appear with a delay of $(0.14 \pm 0.05)$ ps with respect to the initial photoexcitation, which is less than the S$_2$ depopulation time constant. This behavior provides evidence for the ballistic nature of the ring-opening once the S$_1$ state is reached. The resulting biradical species react further within $(1.2 \pm 0.2)$ ps via two rival fragmentation channels yielding ketene and ethylene, or propene and carbon monoxide. Our study showcases both the value of gas-phase ultrafast diffraction studies as an experimental benchmark for nonadiabatic dynamics simulation methods and the limits in the interpretation of such experimental data without comparison to such simulations.
\end{abstract}


\maketitle 

\section{Introduction}
Ultrafast dynamics in the excited states of organic molecules are well-known as key processes in many photochemical reactions. Examples with biological relevance are the photosynthesis reactions in plants\cite{cheng_dynamics_2009} and the primary reaction in the human vision process.\cite{polli_conical_2010} The underlying mechanisms of ultrafast charge and energy flow are still not fully understood due to the molecular complexity, compounded by the involvement of the potential energy surfaces of multiple electronic states. Additionally, population transfer between these states often involves nonadiabatic dynamics through conical intersections, where the Born-Oppenheimer approximation is invalid.\cite{schuurman_dynamics_2018}

Studies of isolated, small organic model systems in the gas phase have emerged as a promising approach to gaining a fundamental understanding of general photochemical reaction mechanisms involving such nonadiabatic dynamics. Due to the absence of an environment and the resulting limited complexity of the system, experimental observables can provide a detailed picture of the investigated process. Such studies also provide the opportunity for a direct comparison with the results of high level quantum chemical simulations of the nonadiabatic dynamics.\cite{curchod_ab_2018} This combination of experimental and simulation approaches is now well-established.\cite{wolf_photochemical_2019, champenois_conformer-specific_2021, liu_rehybridization_2023, liu_ultrafast_2024, chakraborty_time-resolved_2024, figueira_nunes_monitoring_2024, pathak_tracking_2020, borne_ultrafast_2024, stankus_ultrafast_2019, minitti_imaging_2015, liu_spectroscopic_2020, coates_vacuum_2018, neville_substituent_2016, wolf_hexamethylcyclopentadiene_2014, yang_simultaneous_2020, champenois_femtosecond_2023} It has the potential to lead to the development of a predictive understanding of ultrafast photochemical dynamics and the eventual control of reactive outcomes through the development of structure-function relationships as design principles for photochemistry.

The development of such structure-function relationships will require the exploration of large parameter spaces, which is unfeasible by only experiments, given the required effort for each experiment and typical experimental cycles of months and years from their preparation to the understanding and verification of their results. Instead, the majority of the survey of parameter spaces for structure-function relationships could be performed \textit{in silico} by quantum chemical simulations. However, the accuracy of electronic structure methods for excited states is in general lower than for the electronic ground state. Additionally, the specific limits and validity of the large number of available methods for calculating excited state electronic structure and for propagating the nuclear wavepacket on and between excited states are largely underexplored, except for a limited number of studies.\cite{chakraborty_nonadiabatic_2022, chakraborty_time_2021} Therefore, there is a clear need for dedicated efforts to benchmark these simulation approaches.

A key aspect of such benchmark studies is the choice of the experimental observable, which needs to fulfill three requirements: (1) It must provide meaningful information about an important aspect of the investigated ultrafast process, e.g.~ the lifetime of a specific excited state or the timescale of a bond dissociation. (2) It must be interpretable without extensive theory input, to allow for an independent experimental result. (3) It must be accessible by simulations with high fidelity. Otherwise, ambiguities can arise in the case of disagreements between experimental and simulation results, as it is unclear whether the disagreement originates from an artifact in the simulation of the observable or from the underlying excited state dynamics simulation. 

Ultrafast X-ray and electron diffraction \cite{yong_time-resolved_2023, centurion_ultrafast_2022} enabled by X-ray free electron lasers\cite{dunne_free_2023} and novel electron sources\cite{weathersby_mega-electron-volt_2015} are powerful techniques capable of generating valuable new experimental observables.\cite{centurion_ultrafast_2022} (1) Diffraction is selectively sensitive to the motion of the nuclei, providing unambiguous signatures e.g.~from dissociation or isomerization reactions. (2) Diffraction signatures can be intuitively interpreted by transformation into real space. The resulting atomic pair distribution functions (PDFs) directly show relative motion of atoms as changes in distance-dependent pair density. (3) Diffraction can be simply simulated with a high level of fidelity within the independent atom model (IAM).\cite{wolf_photochemical_2019, champenois_conformer-specific_2021, liu_rehybridization_2023, liu_ultrafast_2024} 

The IAM treats the atoms in a molecule as non-interacting, neglecting any distortion of the atomic valence electron distribution from chemical bonding. Therefore, IAM diffraction signatures can be simulated with minimal computational effort using a molecular geometry and tabulated or computed scattering form factors. In cases where the IAM is not accurate enough,\cite{stankus_ultrafast_2019, yong_observation_2020, yang_simultaneous_2020, champenois_femtosecond_2023} powerful approaches have been developed to simulate the diffraction observable from the explicit electron density distribution, which can be obtained from an electronic structure calculation.\cite{parrish_ab_2019, northey_ab_2014} 

\begin{figure}
    \centering
    \includegraphics[trim= 0 0 120 0,clip, width=\linewidth]{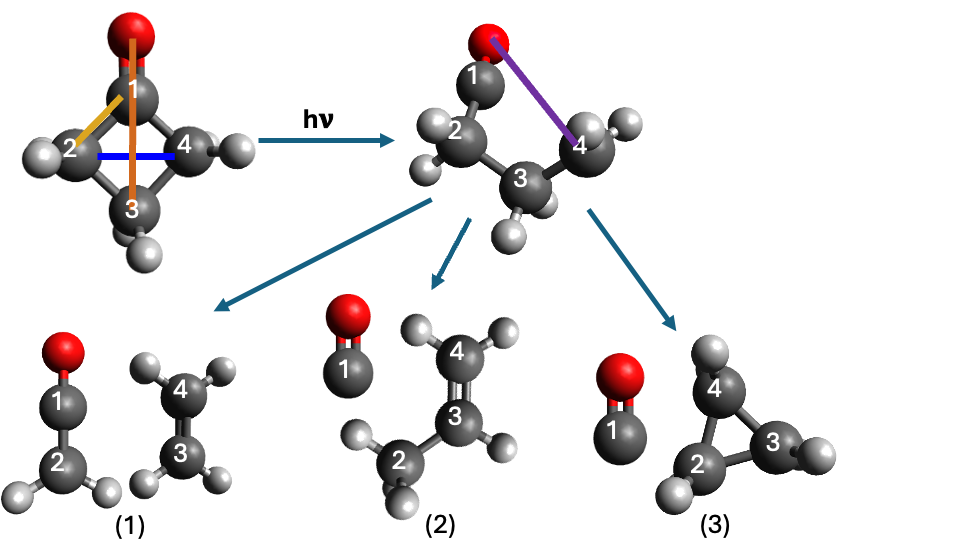}
    \caption{Overview of possible intermediate and fragment structures generated by cyclobutanone photolysis. Cyclobutanone (left) can react via a Norrish Type-I reaction to form a ring-opened biradical product (geometry taken from ref.~\cite{xia_excited-state_2015}). This primary biradical compound can react further via three possible secondary dissociation channels, (1) to ketene and ethylene, (2) carbon monoxide and propene, or to (3) carbon monoxide and cyclopropane.\cite{benson_photochemical_1942, campbell_mechanistic_1967, lee_laser_1977,diau_femtosecond_1998, diau_femtochemistry_2001, kao_effects_2020}  Examples for atomic distances in the first (gold), second (blue), third (brown), and fourth (purple) coordination spheres are shown by color-coded lines. Carbon atoms are shown in dark gray, oxygen atoms in red, and hydrogen atoms in light gray.}
    \label{fig:scheme}
\end{figure}

Therefore, UED is an ideal observable for the present study, which serves as the experimental complement to the simulation studies published in this special issue.\cite{bennett_prediction_2024, eng_photochemistry_2024,hait_prediction_2024, hutton_using_2024, jaiswal_ultrafast_2024, janos_predicting_2024, lawrence_mash_2024, makhov_ultrafast_2024, martin_santa_daria_photofragmentation_2024, miao_casscfmrci_2024, suchan_prediction_2024, miller_ultrafast_2024, mukherjee_prediction_2024, peng_photodissociation_2024, vindel-zandbergen_non-adiabatic_2024} We target the ultrafast photochemistry of organic carbonyls, which are important species in atmospheric chemistry.\cite{akimoto_atmospheric_2016} They exhibit rich photochemistry, including Norrish Type-I reactions leading to the dissociation of a carbon-carbon bond adjacent to the carbonyl group (see Fig.~\ref{fig:scheme}), Norrish Type-II reactions leading to cleavage of a carbon-carbon bond two removed from the carbonyl group via hydrogen atom transfer, and roaming-type reactions.\cite{hause_roaming-mediated_2011, endo_capturing_2020} 

This chemistry is triggered in the atmosphere by photoabsorption into broad, weak bands peaking at $\lambda \approx 280$ nm.\cite{akimoto_atmospheric_2016} The bands correspond to weak transitions to excited states characterized by a single electron transition from a carbonyl oxygen lone pair (n) orbital to the carbonyl $\pi^*$ orbital (n$\pi^*$ character). Their relevance for the tropospheric chemistry of organic carbonyls despite their weak absorption cross-sections comes from the fact that the long wavelength tail overlaps with the short wavelength end of the solar spectrum that reaches the troposphere. However, similar photochemical processes can also be triggered by excitation to Rydberg states at $\lambda \approx 200$ nm with significantly larger absorption cross-section.\cite{campbell_mechanistic_1967, otoole_vacuum-ultraviolet_1991}

The molecule cyclobutanone (CB) has a number of favorable characteristics for use as a model system in a study focusing on the structural dynamics involved in carbonyl photochemistry: (1) It is a comparably small quantum system and is thus accessible to detailed study by contemporary high level quantum chemical methods. (2) Its rigid structure with the strain of a four-membered ring provides a well-defined starting geometry for the reaction. (3) The ring structure prevents the formation of the conformations necessary for a hydrogen atom transfer to the carbonyl oxygen, which is the initial step of the Norrish Type-II reaction. Therefore, the dynamics are constrained to the Norrish Type-I pathway. (4) The Norrish Type-I reaction corresponds to an opening of the four-membered ring, which provides clear and characteristic diffraction signatures. (5) Its vapor pressure is high enough to make it accessible to a broad set of gas-phase time-resolved experimental methods

The CB molecule ($C_4H_6O$) is not completely planar in its ground state. Instead, the carbonyl bond is slightly bent out of the ring-plane reducing the symmetry to C$_s$.\cite{kuhlman_symmetry_2012, tamagawa_molecular_1983, stigliani_structure_1976} The majority of studies of the photochemistry of CB focus on its reactivity after excitation to the weakly absorbing n$\pi^*$ state. Photolysis studies\cite{benson_photochemical_1942, campbell_mechanistic_1967, hemminger_fluorescence_1972, lee_laser_1977} identified three different fragmentation channels (see Fig.~\ref{fig:scheme}) producing ethylene and ketene
\begin{equation}\label{eq:c1}
    C_4H_6O \rightarrow C_2H_4 + CH_2CO
\end{equation}
or carbon monoxide and either propene
\begin{equation}\label{eq:c2}
    C_4H_6O \rightarrow C_3H_6 + CO
\end{equation}
 or cyclopropane
\begin{equation}\label{eq:c3}
    C_4H_6O \rightarrow cC_3H_6 + CO.
\end{equation}
Some of these photolysis studies\cite{campbell_mechanistic_1967, hemminger_fluorescence_1972, lee_laser_1977} suggest  that the propene products (channel \ref{eq:c2}) arise via  acyclization of the cyclopropane products generated in channel \ref{eq:c3}. The photochemistry triggered by n$\pi^*$ excitation was also the subject of a number of time-resolved mass spectrometry studies from the Zewail group\cite{pedersen_validity_1994, diau_femtosecond_1998, diau_femtochemistry_2001} and a recent investigation in the solution phase using transient absorption spectroscopy.\cite{kao_effects_2020} 

The mechanistic picture of the photofragmentation from the photolysis and time-resolved spectroscopy studies is not completely clear, since both fragmentation via a Norrish Type-I reaction\cite{benson_photochemical_1942, campbell_mechanistic_1967, lee_laser_1977,diau_femtosecond_1998, diau_femtochemistry_2001, kao_effects_2020} yielding a biradical intermediate (see Fig.~\ref{fig:scheme}) and direct fragmentation via concerted cleavage of two C-C bonds\cite{pedersen_validity_1994} have been proposed. The two mechanisms could not be unambiguously differentiated by the time-resolved mass-spectrometry studies,\cite{pedersen_validity_1994, diau_femtosecond_1998, diau_femtochemistry_2001} since they were unable to determine if the observed ion fragments derived from neutral products formed before interaction with the photoionizing probe, or from fragmentation in the cations formed by photoionization.\cite{koch_understanding_2015}

Photolysis by excitation of the n3s Rydberg state at $\lambda = 193$ nm\cite{trentelman_193-nm_1990} results in similar fragmentation channels as observed after excitation of the n$\pi^*$ state, suggesting that population in the n3s state can access the n$\pi^*$ state via internal conversion in accord with expectations based on Kasha's rule.\cite{kasha_characterization_1950} The relative yields from the fragmentation channels \ref{eq:c2} and \ref{eq:c3} were measured after photolysis at 200 nm with hydrogen flame detection and determined to be 2.4:1.\cite{campbell_mechanistic_1967}

The ultrafast dynamics of CB after excitation at $\lambda = 200$ nm have been investigated by a combination of time-resolved photoelectron spectroscopy and time-resolved mass spectrometry.\cite{kuhlman_coherent_2012} The time-dependences of the parent ion and a fragment mass corresponding to the ketene and the cyclopropane or propene fragments were fitted to biexponential decays with time constants of 0.08 ps and 0.74 ps for the parent, and  0.12 ps and 0.79 ps for the fragment, respectively. The time-resolved photoelectron spectroscopy study identified a signature displaying a biexponential decay with time constants of 0.31 ps and 0.74 ps, suggesting depopulation of the n3s state on an ultrafast timescale via internal conversion to the n$\pi^*$ state. The decay of the photoelectron signature was observed to be modulated both in position and intensity by an oscillatory component with a period of 0.94 ps, which could be associated with motion along a ring-puckering mode in the n3s state.\cite{kuhlman_pulling_2013}

The evidence for ultrafast nonadiabatic dynamics between states of different electronic character and the rich photochemistry, with several intermediates and products, make CB an ideal benchmark molecule for a challenge for theoreticians to predict non-adiabatic photochemical dynamics in advance of an experiment, as pursued in this special issue.\cite{bennett_prediction_2024, eng_photochemistry_2024,hait_prediction_2024, hutton_using_2024, jaiswal_ultrafast_2024, janos_predicting_2024, lawrence_mash_2024, makhov_ultrafast_2024, martin_santa_daria_photofragmentation_2024, miao_casscfmrci_2024, suchan_prediction_2024, miller_ultrafast_2024, mukherjee_prediction_2024, peng_photodissociation_2024, vindel-zandbergen_non-adiabatic_2024} 

Thus, the purpose of this study is to provide (1) an experimental benchmark to the predictions published in this special issue and (2) information about the mechanism and timescale of the photochemical reaction dynamics of CB following photoexcitation. In the following, we will discuss the results of our UED study of the photochemical dynamics of CB after excitation to the n3s state at $\lambda = 200$ nm. Since our study is the experimental complement to the companion simulation studies, we seek to isolate the analysis of the experimental data as much as possible from the simulation results. Therefore, we will use only a basic level of quantum chemical calculations (geometry optimizations in the ground state at the density functional theory (DFT) level) resulting in an interpretation of the experimental results, which is independent from more extensive simulations.

\section{Methods}\label{sec:methods}

The experiment is carried out at the SLAC mega\-electronvolt ultrafast electron diffraction facility.\cite{weathersby_mega-electron-volt_2015, lin_imaging_2023} The setup for gas-phase ultrafast electron diffraction experiments at the facility is described in detail in Refs.~\cite{lin_imaging_2023, shen_femtosecond_2019}. In short, we separate the output of a 800 nm, 50 fs Ti:Sapphire laser system into two beam paths. The pulses of the probe beam path are frequency tripled and directed onto the photocathode of an S-band radio frequency (RF) gun. They eject an ultrashort pulse containing $\approx 10^4$ electrons, which are subsequently accelerated to 4.3 MeV and focused to a spot size of 200 $\mu$m full width at half maximum (FWHM) in the interaction region of a gas-phase experimental chamber. The pulses of the second, pump beam path are frequency quadrupled to a central wavelength of 200 nm. They are attenuated to a pulse energy of 8.6 $\mu$J, focused using reflective optics, and overlapped with the electron pulses at a 1$^\circ$ angle through an in-vacuum holey mirror into the interaction region of the experimental chamber. The diameter of these pump laser pulses in the interaction region is 490 $\mu$m  x 250 $\mu$m FWHM to over-fill the electron spot. The experimental response function that includes the effects of the optical and electron pulse length as well as the relative arrival time jitter is estimated based on a previous measurement\cite{wolf_photochemical_2019} and the pump pulse duration (130 fs FWHM) to be 190 fs.

Cyclobutanone (purity 99 \%) is purchased from Sigma Millipore and used without further purification. The compound has a vapor pressure of 43 torr at room temperature.\cite{benson_photochemical_1942} This pressure is reduced with the help of a flow controller (MKS) to 1 torr. The resulting low-pressure gas is delivered into a static-filler 3 mm flow cell with 500 $\mu$m diameter entrance and exit holes for the pump and probe beams.\cite{lin_imaging_2023} The experiment is performed at a repetition rate of 360 Hz. Diffracted electrons are detected by a combination of a phosphor screen and an electron multiplying charge-coupled device (EMCCD) camera. Based on the relative levels of static and dynamic signal, we estimate that we excite about 0.3 \% of the molecules (see the Supplementary Information (SI), sect.~S1 for details). Therefore, we do not expect to observe any contributions from multiphoton absorption. Time-dependent diffraction is measured at a series of time delay points between -4.7 ps and 2.1 ps in each scan, where negative delays correspond to the electron probe arriving before the optical pump in the sample. The separation between time delay points is 0.05 ps, except for the earliest and latest delay points where it is considerably larger. At each time delay point, we integrate the diffraction signal for 10 s (3600 shots). The full data set includes 393 such scans of all delays. The processing of the raw diffraction patterns obtained is described in detail in the SI, sect.~S2.

For the simulation of the signatures of the secondary fragmentation products, their geometries and that of CB are optimized using B3LYP/def2-SVP as implemented in ORCA.\cite{neese_software_nodate, neese_efficient_2009, helmich-paris_improved_2021, neese_shark_2023} The geometry of the minimum energy conical intersection between the n$\pi^*$ and ground states from Ref.~\cite{xia_excited-state_2015} is used for the biradical geometry (see the SI, sect.~S3 for further details). These geometries are used to simulate electron diffraction signatures within the IAM using a publicly available code \cite{wolf_diffraction_simulation_nodate} with atomic scattering cross-sections, which are evaluated by the ELSEPA program.\cite{salvat_elsepadirac_2005} Momentum-transfer space $\Delta I/I(s)$ signatures are simulated using 
\begin{equation}
    \Delta I/I \left(s\right) = \frac{I_{P}\left( s\right)-I_{CB}\left(s\right)}{I_{CB}\left(s\right)},
\end{equation}
where $I_P$ and $I_{CB}$ are the simulated IAM diffraction intensities of a specific product and CB, respectively. 

Experimental and simulated PDF and $\Delta$PDF curves are evaluated using an approach which is detailed in the SI, sect.~S4.

\begin{figure*}[ht!]
    \includegraphics[width=\textwidth]{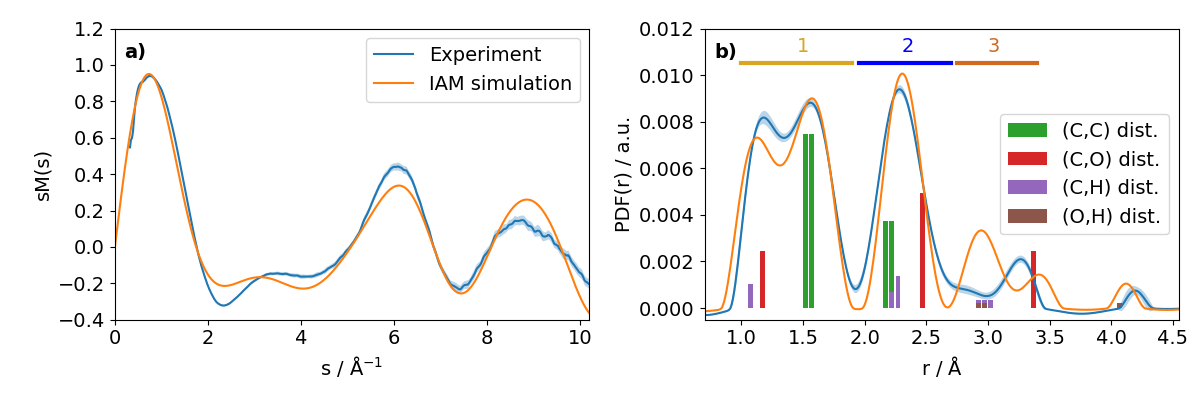}%
    \caption{\label{fig:static}Static structural information from electron diffraction. a) Comparison of the momentum-transfer space (sM($s$)) signature of CB with a simulation within the independent atom model (IAM). b) Atomic pair distribution functions (PDF($r$)) from the real-space transformation of the experimental and IAM signatures of part a). They are compared with histograms of the atomic distance distribution in the ground state equilibrium geometry of CB, which are color-coded with respect to the involved elements and scaled with respect to the atomic scattering cross-sections of the involved atoms. Additionally, the distance ranges of the first three coordination spheres in the PDF are marked by horizontal bars and correspond to color-coded lines in Fig.~\ref{fig:scheme}. Error bars for the experimental plots of parts a) and b) are based on a bootstrapping analysis and are represented by shaded areas.}%
\end{figure*}

\section{Results}
\subsection{Static Structural Information}
Figure \ref{fig:static} shows static structural information from the reactant CB in a) momentum-transfer space as the modified diffraction intensity sM($s$) and b) real space as an atomic pair distribution function PDF($r$). The sM($s$) curve in Fig.~\ref{fig:static} a) is compared to a simulation based on an optimized geometry of CB and the IAM (see sect.~\ref{sec:methods} for details). The experimental sM($s$) curve is truncated below 0.3 \AA$^{-1}$ due to the presence of a hole in the diffraction detector through which the undiffracted electrons pass. The agreement between experimental and simulated sM($s$) is good in terms of the positions of all the maxima and minima. Discrepancies in the amplitude can have contributions from uncompensated deviations in the $s$-dependence of the electron detector response, the finite width of the ground state geometry distribution, and deviations from the IAM.\cite{champenois_femtosecond_2023, yang_simultaneous_2020, bartell_effects_1964}

Figure \ref{fig:static} b) shows the real-space representations of the two curves in Fig.~\ref{fig:static} a) as experimental and simulated PDFs. They are compared with a histogram of the atomic distances in CB which is color-coded and scaled according to the involved elements and their electron scattering cross-sections. The histogram excludes the (H,H) distances due to the small electron scattering cross-section of hydrogen atoms. The atomic distances of the histogram are generally reproduced well by the PDFs.

We will discuss the various (C,C) and (C,O) distances in the histogram within the framework of atomic coordination spheres. The (C-C) and (C-O) bond distances, e.g.~the (C$_1$-C$_2$) distance of CB in Fig.~\ref{fig:scheme}, constitute the first coordination sphere. Signatures from the first coordination sphere appear in the PDFs of Fig.~\ref{fig:static} b) in a range between $r = 1$ \AA~and $r = 2$ \AA, which is marked by a gold bar. Distances between atoms separated by two bonds, e.g.~the (C$_2$,C$_4$) distance of CB in Fig.~\ref{fig:scheme}, constitute the second coordination sphere with values ranging from $r =2$ \AA~to $r=2.7$ \AA~as marked by a blue bar in Fig.~\ref{fig:static} b). The third coordination sphere with distances between atoms separated by three bonds, is comprised of a single distance, (O,C$_3$) in CB in Fig.~\ref{fig:scheme}. Signatures from the third coordination sphere appear beyond $r = 2.7$ \AA~in the PDFs in Fig.~\ref{fig:static} (marked by a light brown bar).

The differences in (C-O) and (C-C) bond distances in the first coordination sphere are resolved in the double-peak structure at $r = 1.2$ \AA~and $r=1.6$ \AA~in both PDFs. The peak at $r=1.2$ \AA~has additional, weak contributions from (C-H) bond distances. The (C,C) and (C,O) distances in the second coordination sphere are too close to be resolved in the PDFs, which results in both cases in a single peak at $r=2.3$ \AA. This peak position value is significantly smaller than in the PDFs of six-membered carbon ring molecules (see e.g.~\cite{champenois_conformer-specific_2021, fnunes_photo-induced_2024}). Therefore, it provides clear evidence for the presence of the four-membered ring structure of CB. 

The area between $r=2.7$ \AA~and $r=3.4$ \AA~containing distances in the third coordination sphere (one (C,O) and several (C,H) and (O,H) distances) shows the largest disagreements between experimental and simulated PDFs. The simulated PDF shows a strong peak in the area of a number of (C,H) and (O,H) distances whereas the experimental PDF only exhibits a shoulder. In turn, the simulated PDF shows a much weaker peak in the area of the (C,O) distance than the experimental PDF. The disagreement can be at least partially explained by a deficiency in the IAM approximation. 
The assumption of atomic behavior made by the IAM is a good approximation for inner-shell electrons, as they are not involved in chemical bonding. Hydrogen atoms lack inner-shell electrons. Thus, when hydrogens are involved in chemical bonds, their electron density distribution exhibits particularly strong deviations from the assumptions by the IAM.\cite{bennani_differential_1979} Therefore, the IAM description overestimates the localization of electron density at the hydrogen atoms, which leads to an overestimation of the amplitude of the (C,H) and (O,H) contributions. 

The small peak beyond $r = 4$ \AA~originates from the two longest (O,H) distances in the molecule. In the following, we will reduce the discussion of the time-dependent structural evolution of photoexcited CB to the stronger signatures of (C,C) and (C,O) distances and neglect the relatively  small contributions from (C,H), (O,H) and (H,H) distances.

\subsection{Signatures of the photoinduced structural dynamics of CB}

\begin{figure*}[ht!]
    \includegraphics[width=0.9\textwidth]{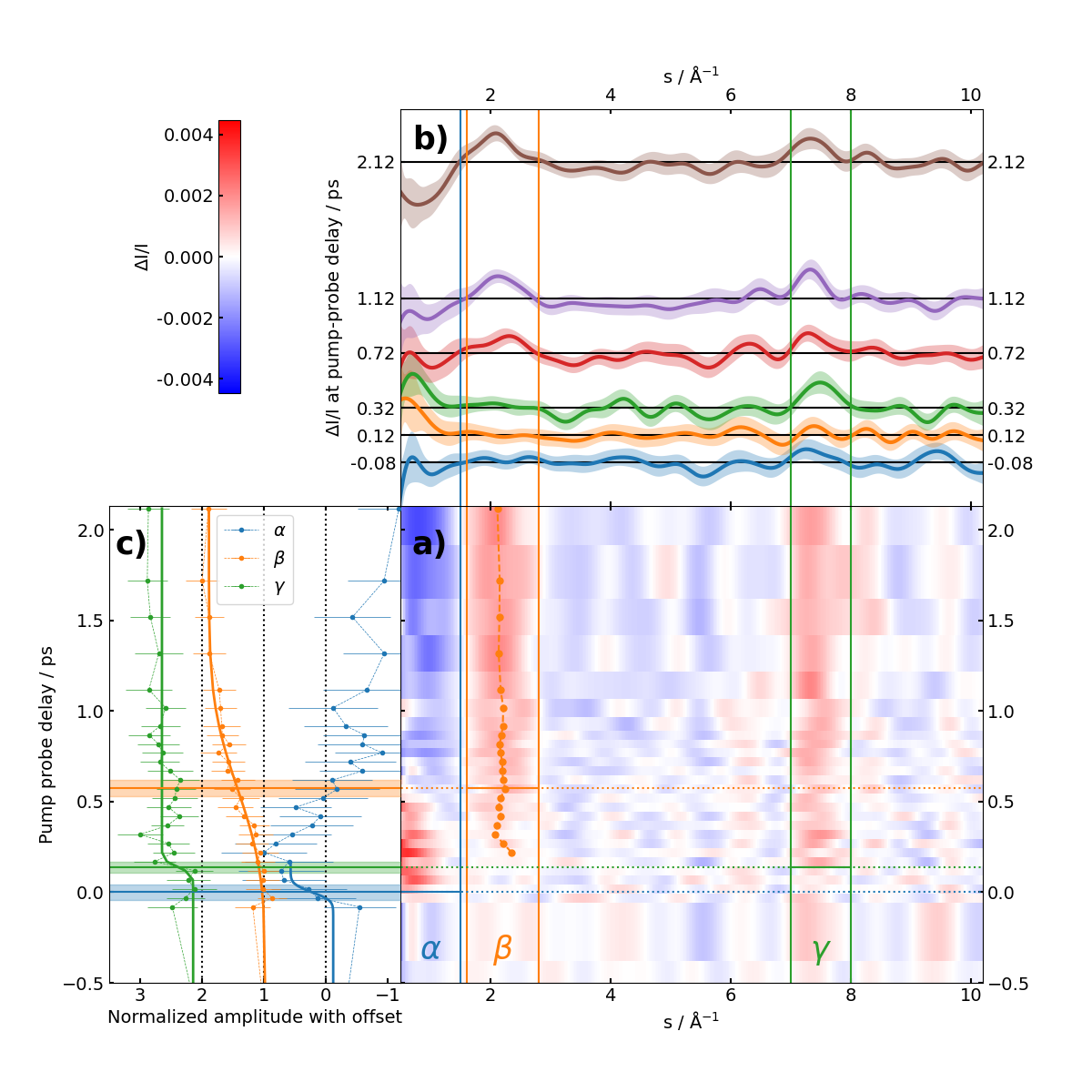}%
    \caption{\label{fig:2d} Diffraction signatures of the UV-induced dynamics. a) False-color map of the signatures represented as $\Delta I/I$. Three different s-ranges with time-dependent changes are marked by vertical lines and labeled as $\alpha$, $\beta$, and $\gamma$. The time-dependent center of mass of the positive contributions in the $\beta$ range are marked by orange dots. b) $\Delta I/I$ signatures at different pump-probe delays. The curves are vertically offset from each other to reflect their delay time. Error bars are based on a bootstrapping analysis and are represented by shaded areas. c) Time-dependent $\Delta I/I$ intensities integrated over the $\alpha$, $\beta$, and $\gamma$ ranges. The experimental data are plotted together with error function fits. The centers of these error functions are marked by horizontal lines, which are also extended into part a). The shaded areas around the horizontal lines represent the uncertainty of the fit. The three plots are horizontally offset from each other. The respective zero levels are marked by dotted lines.}%
\end{figure*}

Figure \ref{fig:2d} shows time-dependent electron diffraction signatures of the structural response of CB to photoexcitation at $\lambda = 200$ nm in momentum transfer space. An overview of the time-dependent signatures in the diffraction signal containing both contributions from the photoexcited population and the corresponding ground state population bleach is given by the false-color plot of the percentage difference diffraction ($\Delta I/I(s)$ in Fig.~\ref{fig:2d} a). We identify four different transient features in the three marked regions of Fig.~\ref{fig:2d} a). The strongest of them is a positive and short-lived feature in the $s$-region below 1.5 \AA$^{-1}$, which is marked by a blue vertical line and labeled as $\alpha$. It is superimposed with a negative signature in the same $s$-range. This signature appears at later delays and persists beyond 2 ps after optical excitation. This behavior is visible in the time-dependent integrated intensity of the $\alpha$ range plotted in Fig.~\ref{fig:2d} c)). We refer to the short-lived positive feature in the $\alpha$ region as $\alpha_1$ and the long-lived negative feature as $\alpha_2$.

The $\alpha_2$ feature is accompanied by a positive feature with similar time-dependence in the range between $s=1.6$ \AA$^{-1}$ and $s=2.8$ \AA$^{-1}$, which is marked by orange vertical lines and labeled as $\beta$. It shows a delayed onset with respect to the $\alpha_1$ feature (see the time-dependence of the integrated $\alpha$ and $\beta$ regions in Fig.~\ref{fig:2d} c)) and shifts slightly to lower $s$-values within the first 0.5 ps after optical excitation. This behavior is illustrated by orange dots in Fig.~\ref{fig:2d} a), which represent the center of mass of the $\beta$ feature. Due to the initial presence of the positive $\alpha_1$ feature, it cannot be assessed if the $\alpha_2$ feature is shifting on the same timescale. The position of the negative $\alpha_2$ feature shifts at about 0.5 ps to smaller $s$-values. However, this shift is likely a result of the superposition of the $\alpha_1$ and $\alpha_2$ contributions.

A fourth delay-dependent feature appears in the range between $s=7$ \AA$^{-1}$ and $s=8$ \AA$^{-1}$, which is marked by green vertical lines and labeled as $\gamma$. It appears close to time zero, earlier than the $\beta$ feature, but with a slight delay with respect to the $\alpha_1$ feature. It remains approximately constant over the investigated delay time window. The delay in the onset with respect to the $\alpha_1$ feature is clearly visible in the lineouts of Fig.~\ref{fig:2d} b): The lineout at 0.12 ps delay already shows a positive signature in the $\alpha$ region, whereas a signature in the $\gamma$ region only starts to appear in the subsequent lineout at 0.32 ps. 

We quantify the delay in the onset of the $\gamma$ feature with respect to $\alpha_1$ feature by fitting error functions to the integrated signal S
\begin{equation}
    S\left(t\right) = A_0 + A \left[1+ erf\left(2\sqrt{2\ln\left(2\right)}\frac{t-t_0}{\tau}\right) \right]
\end{equation}
for the integrated ranges in Fig.~\ref{fig:2d} c). $S\left(t\right)$ is the integrated intensity in the respective regions, $A_0$ accounts for small baseline offsets at negative pump-probe delays, $A$ is a scaling factor, $t_0$ is the center of the error function, and $\tau$ its width. For the $\alpha$ region, only early time delays are included in the fit since the later intensity evolution shows a trend towards negative values which cannot be modeled by the simple fit function. The fitted error function center to the $\gamma$ feature shows a delay with respect to the center of the fit to the $\alpha_1$ feature of (0.14 $\pm$ 0.05) ps. The difference in onset times is further visualized by the color-coded horizontal lines in Fig.~\ref{fig:2d} a), which correspond to the fit centers. The growth of the $\beta$ range is qualitatively different from that of the other two features, which can be quantified by the center and the width of the error function fit. The center shows a delay of $(0.58 \pm 0.06)$ ps with respect to the $\alpha_1$ onset. Its width is $(1.2 \pm 0.2)$ ps compared to $(0.1 \pm 0.1)$ ps for the $\gamma$ range. 

We additionally quantify the decay timescale of the $\alpha_1$ feature by a global biexponential fit resulting in a time constant of (0.29 $\pm$ 0.2) ps (see the SI, sect.~S5), i.e.~with an uncertainty in the range of the instrument response function (190 fs). 

\subsection{Comparison to simulated product signatures}
\begin{figure*}[ht!]
\includegraphics[width=\textwidth]{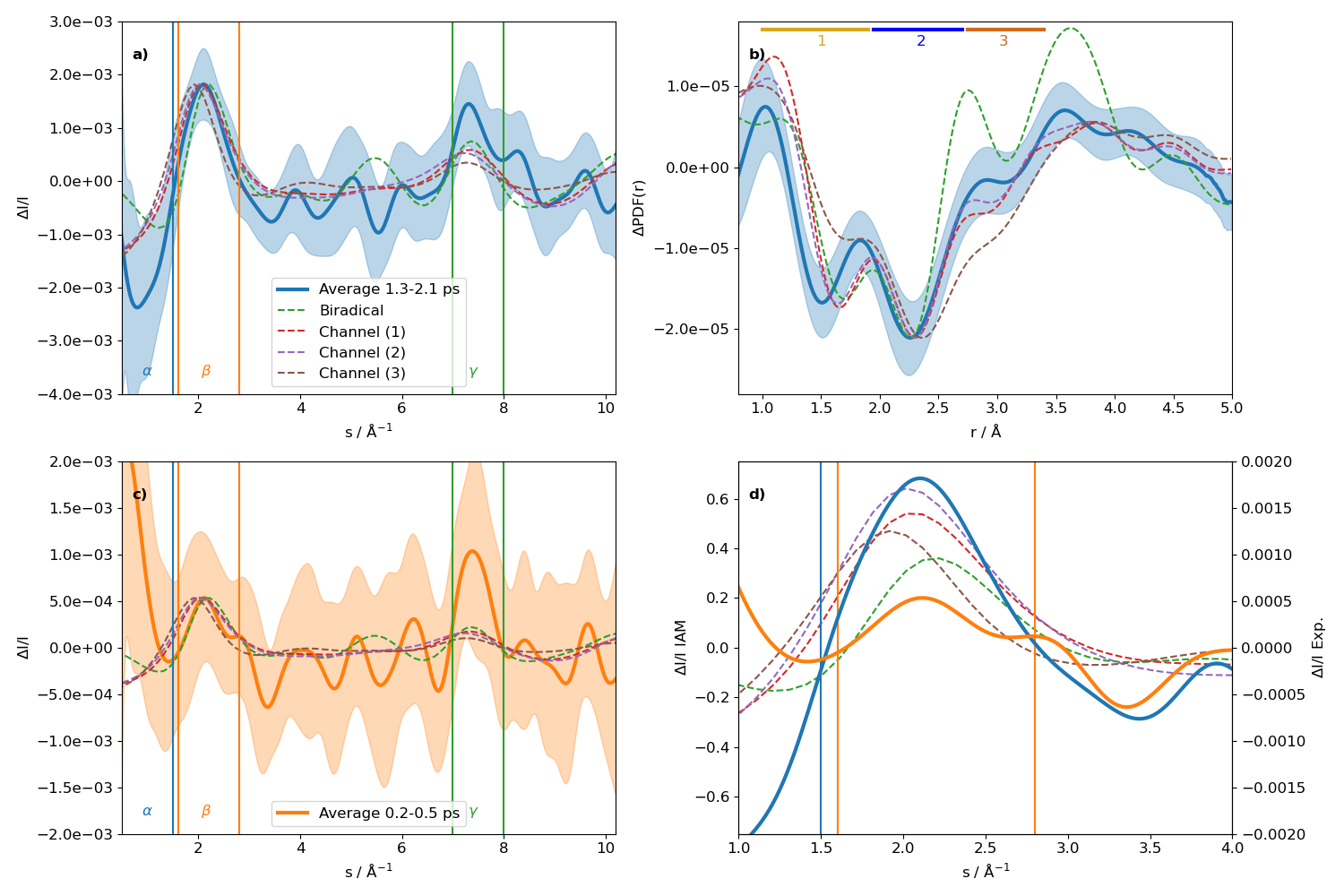}%
\caption{\label{fig:Comp} Comparisons of experimental signatures at different delays with simulations. a) Comparison of the averaged $\Delta I/I$ signature in momentum-transfer space over a delay range of 1.3 ps to 2.1 ps (blue) with independent atom model simulations of the biradical primary product (green) and the three secondary fragmentation channels (red, purple, and brown, see Fig.~\ref{fig:scheme}). The $\alpha$, $\beta$, and $\gamma$ ranges are labeled and marked with vertical, color-coded lines. The simulations are rescaled to match the intensity of the peak in the $\beta$ range of the experimental signal. b) Analogous comparison to a), but after real-space transformation. The simulations are rescaled to match the negative peak at $r=2.3$ \AA~in the experimental difference pair distribution function ($\Delta$PDF). The same horizontal bars as in Fig.~\ref{fig:static} marking the reactant coordination spheres are shown on the top. c) Analogous comparison to a), but with an experimental signature averaged over a delay range of 0.2 ps to 0.5 ps (orange). d) Comparison of peak positions and intensities of the averaged experimental signatures from a) and c) with the simulations. The left y-axis corresponds to the simulations, the right y-axis to the experimental plots.}
\end{figure*}

We start our analysis of the difference diffraction results by comparing an experimental signature averaged across a delay range of 1.3 ps to 2.1 ps, when the difference signal is approximately constant in both intensity and shape, to IAM simulations of static equilibrium product structures in Fig.~\ref{fig:Comp} a) in momentum transfer space. The IAM simulations are based on the geometries of the primary, biradical product of the Norrish type-I reaction and the products of the secondary fragmentation channels (1)-(3), which are visualized in Fig.~\ref{fig:scheme}. They are rescaled to fit the peak intensity of the $\beta$ feature in the experimental signature.

The overall agreement between the experimental data and simulated product signatures is good. In agreement with the experimental results, all IAM simulations show negative contributions in the $\alpha$ region, positive contributions in the $\beta$ region, and positive contributions in the $\gamma$ region. The signatures of all secondary fragmentation channels are rather similar. The biradical simulation differs most from the experimental signal, showing weaker negative contributions in the $\alpha$ region. Additionally, only the biradical has significant positive contributions (at $s=5.5$ \AA$^{-1}$) in the area between the $\beta$ and $\gamma$ regions.

Figure \ref{fig:Comp} b) shows the corresponding comparison between experimental and simulated product signatures in their real-space representation. An atomic distance change during an isomerization reaction such as the primary ring-opening reaction of CB, appears in a $\Delta$PDF($t$) curve as a combination of a negative contribution at the pair distance of the reactant and a positive contribution at the new distance at time $t$. The positive $\Delta$PDF contribution from a distance change can be missing if the distance value is outside ($> 5$ \AA) the considered distance range, e.g. in the case of fragmentation. 

The intense negative contributions of the experimental $\Delta$PDF plot show clear signatures of the dissolution of the four-membered ring: the negative peak at $r=1.5$ \AA~agrees well with typical C-C bond lengths, and is significantly longer than the C-O bond length (1.2 \AA), as discussed previously in relation to Fig.~\ref{fig:static}.  This dissociation of a (C-C) bond in the first coordination sphere must for geometric reasons also affect the $\Delta$PDF in the region of the second (C,C) coordination sphere of the reactant. Accordingly, the $\Delta$PDF plot shows a depletion at $r=2.2$ \AA, i.e.~at $r$ values  significantly smaller than the corresponding second coordination sphere (C,O) distance (2.5 \AA, see Fig.~\ref{fig:static}). Enhancements in the experimental $\Delta$PDF are observed at short  distances ($r<$1.3 \AA) arising from new atomic distances in the photoproducts. The origin of the small positive contributions in the experimental $\Delta$PDF at $r$ values beyond 3 \AA~are further discussed in the SI, sect.~4.2. 

Further information content from the experimental $\Delta$PDF signature can be extracted by comparison with the simulated product $\Delta$PDFs. The $\Delta$PDF signature of the primary biradical product shows qualitatively similar effects of the dissolution of the four-membered ring with negative peaks at $r=1.6$ \AA~and $r=2.2$ \AA~for the same representative structure used above. In contrast to the experimental $\Delta$PDF, however, it also shows two strong positive peaks at $r=2.8$ \AA~and $r=3.7$ \AA. Both peaks can be related to the ring-opening reaction: The 2.8 \AA~peak results from the transformation of a first coordination shell (C-C) distance (of the broken C-C bond) into a third coordination shell (C,C) distance. The peak at $r=3.7$ \AA~results from the transformation of a second coordination shell (C,O) distance into a fourth coordination shell (C,O) distance. The latter is beyond the highest reactant coordination sphere (see color-coded ranges in Fig.~\ref{fig:Comp} b)). The shapes and positions of these positive contributions might be different if a different model than a single geometry is employed for the simulation (see sect.~\ref{sec:prodsigs}), but any ring-opened structure will generate positive contributions in this distance range.

While the simulated $\Delta$PDFs of the fragmentation channels again show the negative fingerprint peaks of the ring dissolution, the two positive peaks of the primary channel are missing. Based on the considerations about positive and negative contributions to $\Delta$PDFs (see above), the corresponding distance values must be beyond $r=5$ \AA. Thus, the missing positive peaks are a direct signature of the fragmentation reaction. An additional, more subtle effect of the fragmentation is evident from the small positive peak at $r=1$ \AA, which is a signature of an overall (C-C) and (C-O) bond distance reduction due to the formation of $\pi$ bonds in some of the secondary products. Since the simulated signatures of the fragmentation product channels fit the experimental $\Delta$PDF substantially better than the signatures of the primary biradical product, specifically in the region between $r=2.5$ \AA~and $r=5$ \AA, we conclude that fragmentation occurs within 1.3 ps after photoexcitation.

Considering the level of agreement of the simulated $\Delta$PDFs of the three possible fragmentation channels with the experimental $\Delta$PDF, we find a high and approximately equal level of agreement for channels 1 and 2. The similarities in the signatures of channels 1 and 2 are not surprising, since the number of distances between carbon and oxygen atoms in each of the coordination spheres are identical: In both cases, two (C-C) bonds are broken leading to negative signatures at $r=1.5$ \AA. The fragmentation also eliminates in both cases two of the second coordination sphere distances and the only third coordination sphere distance in CB providing negative contributions at $r=2.2$ \AA~and $r=3.5$ \AA.

Notably, channel 3 shows a lower level of agreement than channels 1 and 2, specifically around $r=1.5$ \AA~ and $r=3$ \AA. These differences can be traced back to the generation of the cyclopropane fragment, where the loss of the two (C-C) bond distances to the carbonyl carbon is partially compensated by the generation of a new (C-C) bond through the formation of the three-membered ring. In turn, the fragments do not exhibit any distances in the second coordination sphere leading to the differences between $r=2.5$ \AA~ and $r=3.5$ \AA. Based on the agreement of the $\Delta$PDF signatures, we conclude that the distribution of secondary fragments after 1.3 ps is dominated by channels (1) and (2), but cannot exclude minor contributions from channel (3). The $\Delta$PDF signatures of channels (1) and (2) are too similar to make further claims about their relative populations.

Figure \ref{fig:Comp} c) shows a comparison of the same simulations as in Fig.~\ref{fig:Comp} a) with an experimental signal averaged between 0.2 ps and 0.5 ps delay. Again, the simulations are rescaled to fit the peak intensity of the experimental $\beta$ feature. This is the delay range in which a shift of the $\beta$ signature to lower $s$-values is observable (see Fig.~\ref{fig:2d} a)). While the signal-to-noise level of the experimental signature in the $\beta$ region in Fig.~\ref{fig:Comp} is rather poor due to the proximity to the signal onset, the position of the $\beta$ peak is best matched by the simulated biradical signature. This indicates the observation of the ring-opened structure prior to secondary fragmentation. Notably, all simulations qualitatively disagree with the experimental signature in the $\alpha$ region, the origin of which will be discussed in sect.~\ref{sect:dicuss:alpha1}. The presence of the $\alpha_1$ feature prevents a similar analysis in real-space as in Fig.~\ref{fig:Comp} b) (see below). A comparison of the experimental signatures from Fig.~\ref{fig:Comp} a) and c) with the IAM simulations without intensity rescaling in Fig.~\ref{fig:Comp} d) shows that both the experimental signature at early times and the IAM simulation of the biradical geometry have lower intensity (as well as peaks at higher \AA) in the $\beta$ region than the experimental signature at late times and the IAM signatures of the fragmentation channels, respectively, which provides further support for the assignment of the experimental signatures at early times to include contributions from biradical geometries. 

\section{Discussion}
\subsection{Reaction product signatures}\label{sec:prodsigs}
The approximation of photochemical product signatures by $\Delta$PDFs based on IAM simulations from single reactant and product geometries faces known limits. The static PDFs of gas-phase molecules agree with reasonable accuracy with PDFs based on IAM simulations from single geometries, effectively ignoring the geometry blurring effect of vibrational wavefunctions.\cite{hargittai_stereochemical_1988} These effects are strongly increased in the case of photochemical reaction products, where typically a significant amount of the photoabsorbed energy (of 6.2 eV in the present case) is redistributed into the vibrational degrees of freedom of the molecules. Due to its lack of rigidity, the actual $\Delta$PDF signature of the primary, biradical photoproduct can be expected to show the largest deviations from the approximation by a single geometry. The employed geometry for the IAM simulations is in fact not a minimum geometry of the biradical photoproduct, but a minimum energy conical intersection geometry from Ref.~\cite{xia_excited-state_2015} for the conical intersection between the n$\pi^*$ state and the ground state of CB. This geometry resembles more closely a point on the (C$_1$-C$_2$) dissociation coordinate than the biradical minimum geometry (see the SI, sect.~S3). However, the IAM signatures from the minimum geometry lead to qualitatively similar conclusions (see the SI, sect.~S3). Nevertheless, previous studies of ring-opening reactions demonstrate the validity of comparisons with product signatures for qualitative assessments.\cite{wolf_photochemical_2019, champenois_conformer-specific_2021, liu_rehybridization_2023} However, this is an area where more advanced theoretical models of the photodynamics could lead to more detailed insight. 

The fragmentation product geometries from secondary channels (1)-(3) are much more rigid. Additionally, the secondary reactions allow the molecular system to transform a part of the photoabsorbed excess energy from the interatomic degrees of freedom into translational and rotational energy of the fragments. Therefore, we expect the approximation of the IAM signatures of the secondary channel products to be of substantially higher quality than the primary reaction product. Based on these considerations, we are confident about our assignment of the experimental $\Delta$PDF signatures after 1.3 ps to fragmentation channels (1) and (2). 

\subsection{The origin of the $\alpha_1$ feature and the onset of the structural dynamics}\label{sect:dicuss:alpha1}
Similar features like the short-lived $\alpha_1$ feature have been previously observed at low $s$-values.\cite{champenois_femtosecond_2023, yang_simultaneous_2020} In both cases, they could be explained based on \textit{ab initio} scattering simulations as originating from the change in the electronic structure induced by the photoexcitation. Thus, the features cannot be described by the IAM. Previous \textit{ab initio} scattering simulations suggest that they exhibit significant contributions from inelastic electron scattering, which does not contain structural information due to its incoherent nature and is known to appear concentrated at low $s$-values.\cite{duguet_high_1983} The qualitative mismatch between the experimental and simulated signatures at early delays in the $\alpha$ region (Fig.~\ref{fig:Comp} c)) is additional evidence in favor of an assignment of the $\alpha_1$ feature to the non-IAM signature of an electronic excitation in CB. 

These non-IAM signatures were in the two previous studies attributed to states with n3s and n$\pi^*$ characters. The previous findings suggest that we observe population dynamics in one or several electronic states with n-hole character in our present study. Two electronic states of CB with n-hole character exist in the energy range of current interest, the initially excited n3s Rydberg state and a lower-lying n$\pi^*$ state. The latter is likely populated from the n3s state through nonadiabatic dynamics on the sub-ps timescale.\cite{kuhlman_coherent_2012} The relative timing of the onset of the $\alpha_1$ feature, which precedes those of the other observed features, suggests that it at least initially corresponds to population of the n3s state by photoexcitation. Since it originates from a change in the electronic wavefunction of the molecule induced by the photoexcitation, its onset is quasi-instantaneous at time zero. In contrast, signatures from structural dynamics typically experience some delay with respect to the optical excitation, since the nuclei need to move a sufficient amount to generate a difference diffraction signal. Thus, the onset of the $\alpha_1$ feature provides an unambiguous signature of time zero.

Without more advanced simulation of the observable, we cannot exclude the possibility that the $\alpha_1$ feature also contains contributions from population in the n$\pi^*$ state at later delays. However, the fitted decay time constant (0.29 ps, more details in SI sect.~S5) of the $\alpha_1$ feature agrees very well with the faster of the two time constants of the biexponential fit (0.31 ps and 0.74 ps) to the n3s feature in the time-resolved photoelectron spectra from Ref.~\cite{kuhlman_coherent_2012}. Therefore, the $\alpha_1$ feature is likely predominantly caused by population in the n3s state.

Since inelastic electron scattering signatures are strongly localized at small $s$-values, it is highly unlikely that the $\gamma$ region is a non-IAM signature. Additionally, the IAM simulations of primary and secondary reaction products all show positive peaks in the $\gamma$ region. Therefore, we interpret the onset of the signal in the $\gamma$ region as the onset of the structural dynamics. Accordingly, the delay between the $\alpha_1$ and $\gamma$ signatures of 0.14 $\pm$ 0.05 ps is the delay between optical excitation and the onset of significant structural dynamics.

\subsection{Delay and shift in the $\beta$ feature}
The onset of the $\gamma$ feature does not give evidence for the specific nature of the structural change, since it is shared by both the primary and secondary products. The slower onset of the $\beta$ feature and its shift to lower $s$-values provide more conclusive insights into this issue. The signature of the biradical geometry shows a positive contribution to higher $s$-values with respect to the secondary products in the beta region. Thus, a transformation from the biradical to the secondary products would result in the observed shift in the $\beta$ feature to lower $s$-values. Hence, the experimental signature at delays between 0.2 ps and 0.5 ps in Fig.~\ref{fig:Comp} c) can be assigned to the biradical photoproduct. This assignment could be more obvious after real-space transformation. However, the $\alpha_1$ feature dominates the $\Delta I/I$ signal at these early delays. Due to its suspected inelastic scattering nature, a real-space transformation would not yield meaningful results.

One would intuitively expect to observe the onset of the $\beta$ feature to be co-timed with the onset of the $\gamma$ feature at slightly higher $s$-values. Instead, it slowly increases in intensity while also shifting in $s$. This behavior can be explained by considering the much less well-defined geometry (see sect.~\ref{sec:prodsigs}) of the primary photoproduct compared to the secondary products. In the comparison of the $\Delta I/I$ signatures of different possible biradical geometries (see the SI, sect.~S3), the peak positions in the $\alpha$ and $\beta$ regions show the largest variance, while the peak position in the $\gamma$ region is much more stable. Based on this observation, we expect that the experimental signature of the primary photoproduct is much less pronounced than the secondary products in the $\beta$ region, which could explain the slow onset of the $\beta$ feature. Additionally, due to its transitory nature, its population at all but the early pump-probe delays can be expected to be low contributing to the observed weakness in the $\beta$ region.

\subsection{Proposed reaction mechanism based on the UED results}
All above findings and considerations  lead us to the following mechanism for the photochemical reaction dynamics of CB within the first 2 ps after photoexcitation at $\lambda=200$ nm: The photoexcitation populates an n3s Rydberg state which can be followed in the onset of the $\alpha_1$ feature. The n3s state is largely depopulated within 0.29 ps via internal convsersion to the n$\pi^*$ state, which can be observed with the decay of the $\alpha_1$ feature. 

The molecular structure responds within 0.14 $\pm$ 0.05 ps by a Norrish Type-I reaction opening the four-membered ring and resulting in the biradical primary product. The timescale for the onset of this ring-opening reaction is set by the delay of the $\gamma$ feature with respect to the $\alpha_1$ feature. The value of the delay is below the instrument response function of the experiment. However, it occurs between two features in clearly separated s-regions, which makes it well-observable.

The lifetime of the $\alpha_1$ feature (0.29 $\pm$ 0.2 ps) is larger than the difference in onset between the $\alpha_1$ and the $\gamma$ features. However, according to previous time-resolved photoelectron spectroscopy results, the main structural motion enabling depopulation of the n3s state is ring-puckering,\cite{kuhlman_pulling_2013} which can be expected to provide a negligibly weak signature in the difference diffraction patterns. Thus, having traversed the n3s/n$\pi^*$ conical intersection, molecules on the n$\pi^*$ potential must undergo quasi-instantaneous (C$_1$-C$_4$) bond extension and access the conical intersection of the n$\pi^*$ state with the ground state. Accordingly, the diffraction patterns show the onset of \textit{structural} dynamics in the n$\pi^*$ state (onset of the $\gamma$ feature) while the \textit{population} dynamics out of the n3s state (decay of the $\alpha_1$ feature) are still ongoing. Internal conversion through the conical intersection with the ground state results in vibrationally hot ground state molecules.

According to Ref.\cite{xia_excited-state_2015}, the Norrish Type-I reaction coordinate exhibits an energy barrier of 0.29 eV relative to the potential minimum of the n$\pi^*$ state. We can estimate the amount of energy available in the vibrational degrees of freedom of CB after internal conversion from the n3s state based on the excitation photon energy and the position of the n$\pi^*$ transition in its absorption spectrum ($\lambda=280$ nm) to be $\approx$ 1.8 eV. Since this value is substantially higher than the barrier, the barrier is unlikely to play a significant role in slowing down the reaction. Thus, the topology of the n$\pi^*$ potential energy surface is consistent with quasi-instantaneous (C-C) bond extension (ring-opening) towards the conical intersection with the ground state. Given the reported minimum conical intersection (ref.~\cite{xia_excited-state_2015}), we can estimate that the vibrational energy content within the dissociating molecules will have increased to $\approx2.7$ eV at this point. 

Reference \cite{xia_excited-state_2015} suggests that the ring-opening reaction is accompanied by internal conversion through a conical intersection between the $n\pi^*$ state and the ground state. Thus, in the case of a quasi-instantaneous ring-opening reaction after reaching the $n\pi^*$ state, the relative population in the $n\pi^*$ state would be small and accordingly would have negligible contributions to the $\alpha_1$ feature. Previous studies also proposed reaction mechanisms involving intersystem crossing to the triplet manifold.\cite{diau_femtochemistry_2001, trentelman_193-nm_1990} Since our observable is not directly sensitive to the multiplicity of the populated electronic states, we cannot definitively rule out the presence of reaction pathways involving triplet states. However, the observed timescales, which would be extremely fast for intersystem crossing, suggest a reaction mechanism in the singlet manifold.

Kuhlman \textit{et al.}\cite{kuhlman_coherent_2012} observed a biexponential decay of the n3s state feature in their time-resolved photoelectron spectra. We would not expect to observe the same decay of the n3s state feature ($\alpha_1$) due to a superposition with negative contributions from the ring-opening and fragmentation signatures ($\alpha_2$). Since the superposition occurs between a feature with sensitivity to population dynamics and one with sensitivity to structural dynamics, it is difficult to disentangle them in a meaningful way.

The primary ring-opened photoproduct reacts further by fragmentation into secondary product channels (1) and (2) yielding ketene and ethylene and propene and CO, respectively. The dominance of channel (2) over channel (3) is consistent with a photolysis study after 200 nm excitation, which deduced a ratio between channels (2) and (3) of 2.4:1.\cite{campbell_mechanistic_1967} The same study showed evidence that after excitation to the $n\pi^*$ state the formation of the channel (2) products was the result of a tertiary isomerization reaction from channel (3). However, the substantially higher level of vibrational energy available to the secondary fragmentation reaction after excitation to the n3s state, might well lead to a different final product distribution than that observed after $n\pi^*$ excitation.

\section{Conclusion}

In our investigation of the ultrafast structural dynamics of cyclobutanone, without advanced theoretical input we observe the depopulation of the photoexcited n3s Rydberg state within 0.29 ps towards the n$\pi^*$ state, followed by rapid ring-opening via a Norrish Type-I reaction. The primary biradical ring-opening product further fragments within the investigated time window mainly into two secondary product channels creating ketene and ethylene, or propene and CO. The complementary sensitivity of ultrafast electron diffraction to electronic structure changes and nuclear structure changes is instrumental in assigning the reaction mechanism. The observation of easily separable signatures of both electronic state population dynamics and structural dynamics provides rich information content for comparison to simulations of the nonadiabatic dynamics. Additionally, our study emphasizes the limits to the accurate interpretation of experimental data from gas-phase UED without the comparison to trajectory simulations.

\section{Acknowledgement}
This work was primarily supported by the AMOS program within the U.S. Department of Energy Office of Science, Basic Energy Sciences, Chemical Sciences, Geosciences, and Biosciences Division. A.G.~was supported  by the European Union, through Horizon Europe project 123-CO: 101067645. Views and opinions expressed are however those of the authors only and do not necessarily reflect those of the European Union. Neither the European Union nor the granting authority can be held responsible for them. NG, LH and PMW were funded by the US Department of Energy under grant no. DE-SC0017995. Funding for the collaboration including PMW and MC, which also supported SWC and SKS, was provided by the U.S. Department of Energy, Office of Science, Basic Energy Sciences, under award number DE-SC0020276. AJOE is grateful for support from EPSRC grant EP/V026690/1. JS, HVSL, DR, and AR were supported by the Chemical Sciences, Geosciences, and Biosciences Division, Office of Basic Energy Sciences, Office of Science, US Department of Energy, grant no. DE-FG02-86ER13491. The experiment was performed at SLAC MeV-UED, which is supported in part by the Department of Energy Basic Energy Sciences’ Scientific User Facility Division Accelerator and Detector research and development program, the LCLS Facility, and SLAC under Contract Nos. DE-AC02-05-CH11231 and DEAC02-76SF00515.

\bibliography{references}

\end{document}